\documentclass[aps,pre,showpacs,twocolumn,superscriptaddress]{revtex4}
\usepackage{graphicx,color,epsfig}
\usepackage{amssymb,amsmath}

\begin{document}

\title{A mesoscopic model for binary fluids}
 \author{C. Echeverria}
 \affiliation{CeSiMo, Universidad de Los Andes, M\'erida 5251, M\'erida, Venezuela.}
  \author{K. Tucci}
 \affiliation{CeSiMo, Universidad de Los Andes, M\'erida 5251, M\'erida, Venezuela.}
  \affiliation{Grupo de Caos y Sistemas Complejos, Centro de F\'isica Fundamental, Universidad de Los Andes, M\'erida, Venezuela}
 \author{O. Alvarez-Llamoza}
\affiliation{Departamento de F\'isica, FACYT, Universidad de Carabobo, Valencia, Venezuela.}
 \affiliation{Facultad de Ingenier\'ia, Universidad Cat\'olica de Cuenca, Ecuador}
 \author{E. E. Orozco-Guill\'en}
 \affiliation{Programa Acad\'emico de Ingenier\'ia en  Energ\'ia, Universidad Polit\'ecnica de Sinaloa, 82199 Mazatl\'an, Mexico.}
 \author{M. Morales}
 \affiliation{Programa Acad\'emico de Ingenier\'ia en Nanotecnolog\'ia, Universidad Polit\'ecnica de Sinaloa, 82199 Mazatl\'an, Mexico.}
 \author{M. G. Cosenza}
\affiliation{Grupo de Caos y Sistemas Complejos, Centro de F\'isica Fundamental, Universidad de Los Andes, M\'erida, Venezuela}

\begin{abstract}
We propose a model to study symmetric binary fluids, based in
the mesoscopic molecular simulation technique known as multiparticle
collision, where space and state variables are continuous while time is discrete. 
We include a repulsion rule to simulate segregation processes that does not require the calculation of the interaction forces between
particles, thus allowing the description of binary fluids at a mesoscopic scale. 
The model is conceptually simple, computationally efficient, maintains Galilean invariance, and conserves the
mass and the energy in the system at micro and macro scales; while  momentum is conserved globally.
For a wide range of temperatures and densities, the model yields results in good agreement
with the known properties of binary fluids, such as density profile,
width of the interface, phase separation and phase growth.
We also apply the model to study binary fluids in crowded environments with consistent results.
\end{abstract}

\pacs{89.75.Fb, 87.23.Ge, 05.50.+q}

\maketitle

\section{Introduction}

In recent years there has been much interest in the development of
computational models for simulation of fluid dynamics based on particle
interactions \cite{MCG2003,PTBLW2003,mills2013,saunders2013}. In many problems of fluid simulation, the potential
energy of a moderate number of particles is enough to represent some
macroscopic behaviors. However, to study properties such as mobility of colloids,
chemical reactions of macromolecules, fluid diffusion in crowded media, or
dynamics of phase segregation, a large number of particles is required to
obtain good descriptions. For such problems, several techniques of mesoscopic
simulation have been implemented; for example, lattice gas automata \cite{FHP86},
lattice Boltzmann method \cite{Succi01}, dissipative particle dynamics
\cite{HK92,GW97}, smoothed particle dynamics \cite{ER03} and multiparticle
collision dynamics \cite{MK1,MK2,MK3}. Each of these techniques provides a
coarse-grained approach that incorporates conservation laws and the essential
physics while omitting corpuscular details.

Multiparticle collision dynamics, also known as stochastic
rotation dynamics \cite{GIKW09}, is a particle-based technique for complex fluids that
includes thermal fluctuations and hydrodynamic interactions \cite{MK1}. Multiparticle collision dynamics 
has proven to be capable of simulating many soft-matter systems, including
colloid dynamics \cite{MK2,PL04,Hea06}, polymer and proteins dynamics \cite{MRWG05,
RWG07,EK10,ETMK11,EK12,EK14}, vesicles \cite{NG05} and reactive systems \cite{rohlf2008,TK1}. 
Multiparticle collision dynamics has also been employed
to investigate the properties of chemical reaction in crowded environments;
i.e., media containing obstacles \cite{TK2,ETK07,EK1}.

In particular, due to their spatial and temporal scales, binary fluid
systems are susceptible to be simulated through multiparticle collision techniques. There are
two main approaches to simulate a binary fluid in the context of multiparticle collision dynamics.
The first one, proposed by Hashimoto et al. \cite{HCO00}, incorporates an additional
collision step in the multiparticle collision scheme to guide the mean particle flow of each
species in the direction of its density gradient. The authors studied
segregation phenomena in a binary fluid and observed the formation of drop-shape
domains with curvatures that can be described by Laplace's law. An
extension of this approach has been used to describe amphiphilic fluids \cite{ICO04} and
compounds consisting of hydrophilic and hydrophobic parts \cite{SCO02}. Although
this extension of the multiparticle collision technique conserves energy and momentum, it has
not been proven that it leads to thermodynamically consistent results \cite{Drube13}.

The second approach extends multiparticle collision dynamics to a binary mixture where
collisions between particles of different species occur in supercells while the
rest of the multiparticle collision process is carried out in smaller cells \cite{TPIK07}. This approach can 
simulate phase separation phenomena, although with the use of a shifting technique
to ensure Galilean invariance \cite{IK1}. However, phase separation is achieved at very
low temperatures that usually introduce strong correlations among particles.

In this article we propose a multiparticle collision model with a repulsion rule to investigate phase separation processes in a binary fluid
in both free and crowded environments. 
In Section~II we present the  model and introduce the 
repulsion rule for the collision dynamics between the centers of mass of particles from different species. 
We show that this rule keeps Galilean invariance and preserves 
the mass and the energy of the system, and discuss the conservation of momentum.  
In Sec.~III the model is employed to study the behavior of a binary fluid in a free environment. 
We investigate several phenomena in a wide range of temperatures, including the stability of an interface front between the two species, 
phase separation, and phase growth, and calculate the characteristic parameters for such processes. 
The simulation results are shown to be in good agreement with theoretical models.
In Sec.~IV we simulate the properties of a binary fluid in a crowded environment by considering 
 particles of both species moving through a random
distribution of stationary obstacles. Under this conditions,
we describe the stabilization of the interface and the formation of
domains. Conclusions are given in Section~V.

\section{Segregation rules in  multiparticle collision dynamics}

Multiparticle collision models simplify the dynamical description while retaining the essential features 
of molecular dynamics\cite{MK1,MK2,MK3}. We consider a fluid consisting of
particles of two species, $A$ and $B$. The masses of particles of species $A$ and $B$
are $m^A$ and $m^B$, respectively. 
Particles of both species,  with continuous positions and
velocities, free stream between multiparticle collision events
that occur at discrete times $\tau$. To carry out collisions, the
volume ${\cal V}$ of the system is divided into cubic cells with length $\ell=1$ 
labeled by an index $\xi$. 
We denote by $n_{\xi}^\gamma$ the number of particles of species $\gamma$ in the cell $\xi$, where $\gamma$
can take the values $A$ or $B$. The velocity of the center of mass of particles of species $\gamma$ in the cell $\xi$
before collision, is given by
\begin{equation}
\label{eq1}
{\bf  V}_{\xi}^{\gamma} = \frac{1}{n_{\xi}^{\gamma}} \sum_{i=1}^{n_\xi^\gamma} {\bf v}(i)_\xi^\gamma ,
\end{equation}
where ${\bf v}(i)_\xi^\gamma$ is the pre-collision velocity of particle $i$ of species $\gamma$ in the cell $\xi$. 

Segregation in binary fluids can be simulated by including repulsion
effects among species, similar to those employed in models for spinodal
decomposition with molecular dynamics~\cite{LTM1}. Thus, we define  
the center-of-mass velocity of particles of
species $\gamma$ after the all-species collision as,
\begin{equation}
\label{eq2}
{\bf \widetilde V}_{\xi}^{\gamma} = 
\frac{\kappa \rho_{\xi}^{\gamma^*}m_{\gamma^*} \widehat{\bf r}_{\gamma\gamma^*} + {\bf V}_{\xi}^{\gamma}}
{|\kappa \rho_{\xi}^{\gamma^*}m_{\gamma*} \widehat{\bf r}_{\gamma\gamma*} + {\bf V}_{\xi}^{\gamma}|} |{\bf V}_{\xi}^{\gamma}|, 
\end{equation}
where $\gamma^*$ represents a species different from $\gamma$,
$\rho_{\xi}^{\gamma^*}$ is the density of particles of species $\gamma^*$
in the cell $\xi$,
$\widehat{\mathbf r}_{\gamma\gamma^*}$ is the unit vector in the direction 
between the center of mass of species $\gamma$ and $\gamma^*$,
$|\cdots|$ is the vector norm and $\kappa$ is a parameter representing the repulsion force between different species. 
Note that, if $\rho_{\xi}^{\gamma^*}=0$, i.e., if there are no particles of species $\gamma^*$ in the cell $\xi$, 
then the velocity of the  center-of-mass of particles of species $\gamma$ does not change; i.e., 
${\bf \widetilde V}_{\xi}^{\gamma}={\bf V}_{\xi}^{\gamma}$.
We calculate the velocity of particles of species $\gamma$ with respect to the
velocity of the center of mass after the collision as
\begin{equation}
\label{eq3}
 {\bf \widetilde v}(i)_\xi^{\gamma}= {\bf \widetilde V}_{\xi}^{\gamma} + ( {\bf v}(i)_\xi^{\gamma} - {\bf V}_{\xi}^{\gamma}).
\end{equation}

Next, we apply the one-species rotation defined by 
\begin{equation}
\label{eq4}
{\bf v'}(i)_\xi^{\gamma}= \sum_\gamma \left( 
{\bf {\widetilde V}}_\xi^{\gamma} 
+ \widehat{\omega}_\xi^\gamma ({\bf \widetilde v}(i)_\xi^{\gamma} - {\bf \widetilde V}_\xi^{\gamma})
\right) ,
\end{equation}
where $\widehat{\omega}_\xi^\gamma$ is a random chosen rotation operator
applied only to particles of species $\gamma$.

Another way to see the rule introduced in Eq.~(\ref{eq2})
is as a rotation of the velocity vector of the center of mass of each
species in the opposite direction to the center of mass of the other
species. Note that the multiparticle collision  model with repulsion conserves both energy and mass in each
cell $\xi$ after each multiparticle collision event. Linear momentum is conserved
 within a homogeneous phase, but not at the interfaces.

\section{Binary fluid in free environment}
The phenomena of phase separation and  formation of stable interfaces in binary fluids
usually occur at low temperatures.
To study the behavior of the multiparticle-collision repulsion model at low temperatures, we first consider
the case of having a single species. Then, only rules (\ref{eq1}) and (\ref{eq4}) apply, 
with ${\bf \widetilde V}_{\xi}^{\gamma}={\bf V}_{\xi}^{\gamma}$.
The volume is defined as
${{\cal V} = L_x\times L_y \times L_z} = 50^3$, with periodic boundary conditions.
In each simulation step, the rotation operators $\widehat\omega_\xi$ are taken to describe rotations $\pm \pi/2$ about randomly chosen axes.
The number of particles in the system is $N = {\cal V}  \rho$, where $\rho$ is the mean density of particles. 

Assuming that there is no correlation between the events of collision, it has been shown that
the diffusion coefficient for a single species in multiparticle collision dynamics can be approximated by the expression \cite{TK1}
\begin{equation}
\label{equ3}
D = \frac{k_B\,T}{2\,m} \left( \frac{2\rho + 1
      - e^{-\rho}}{\rho - 1 + e^{-\rho}}\right) ,
\end{equation}
where  $k_B$ is the Boltzmann constant and the temperature $T$ is given in reduced units.
This equation is satisfied when the average displacement of the particles between collisions is of the order
of the size of the cell; i.e. when $T \approx  1$.

\begin{figure}[h]
\centerline{\includegraphics[scale=0.25,angle=90]{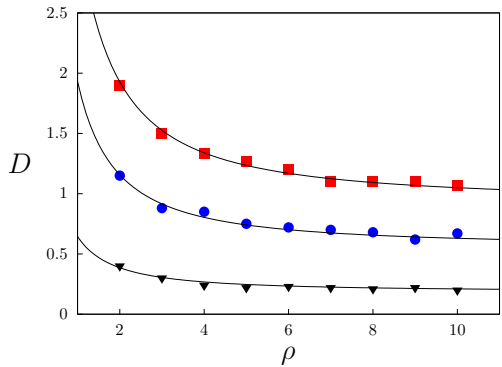}}
\caption{Diffusion coefficient $D$ as a
function of the density of particles $\rho$ for a single species, calculated from simulation of the model for three different 
temperatures:  $T=0.3$ (squares),  $T=0.18$ (circles) and 
$T=0.06$ (triangles). Solid lines correspond to Eq.~(\ref{equ3}).}
\label{fig:D-vs-rho}
\end{figure}

Figure \ref{fig:D-vs-rho} shows the diffusion coefficient $D$ for a single species
as a function of the density, calculated from the simulations and compared with Eq.~(\ref{equ3}), for different temperatures.

Note that at low temperatures the diffusion coefficient calculated
from simulations agrees with the mesoscopic diffusion coefficient given by Eq.~(\ref{equ3}). 
These results show that  multiparticle collision  models can be used to simulate systems with relatively low temperatures ($T=0.06$),
keeping a good diffusive behavior.

To implement the multiparticle-collision model with repulsion, Eqs.~(\ref{eq1})-(\ref{eq4}),
we consider a three-dimensional film with length $L_x =100$ units along $x$, width
$L_y = 100$ units along $y$, and height
$L_z = 2$ units along $z$. We impose periodic boundary conditions
in the $y$ and $z$ directions, and bounce-back reflection boundary conditions on both
the left and the right side of the film along the $x$ direction. The average number of particles per cell 
is assumed to be the same for both species; i. e., $n^A=n^B \equiv n$.
Additionally, we assume species with equal masses; i. e., $m^A=m^B\equiv m$.

First, we study the properties of the interface between the two fluids. 
As initial condition, particles of species $A$ are  uniformly distributed at random
on the right side of the film, so that their mean density, averaged over cells, is
$\rho^A(x \ge 50) = 2 m n$ and $\rho^A(x < 50) = 0$; while particles of species $B$ are similarly distributed 
on the left side of the film; i.e., $\rho^B(x \le 50) = 2 m n$ and $\rho^B(x > 50) = 0$. 

\begin{figure}[h]
\centerline{\includegraphics[scale=0.36,angle=90]{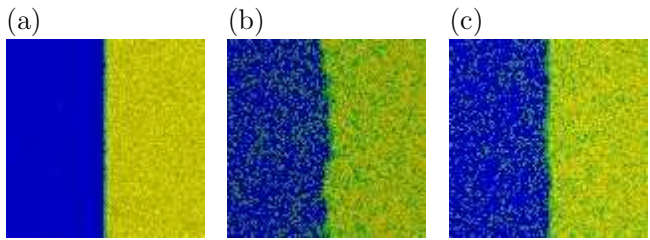}} 
\caption{Snapshots along the $z$-axis of the system for $T=0.12$ and $\kappa=5$.  
Particles of species $A$ are assigned a yellow (light gray) color and  particles of species $B$ are marked in blue (black). 
Green (dark gray) color indicates
the presence of particles of both species. The color intensity is proportional to the density of particles. 
(a) Initial state. (b) State after $t = 10^5$ iterations for $n = 5$. (c) State after $t = 10^5$ iterations for $n=8$.}
\label{fig:snapshot_I}
\end{figure}

Figure~\ref{fig:snapshot_I}(a) shows a snapshot of the 
initial condition of the system. Figures~\ref{fig:snapshot_I}(b) and \ref{fig:snapshot_I}(c) 
show snapshots of the system at $t=10^5$ iterations with parameters $n=5$ and $n=8$, respectively.
Note that in both, Fig.~\ref{fig:snapshot_I}(b) and
Fig.~\ref{fig:snapshot_I}(c), the system
maintains two  phases separated by a thin region where species $A$ and $B$ are mixed. 
That is, the multiparticle-collision repulsion model is able to stabilize the interface for some values of the parameters of the system.

To characterize the interface, we calculate the normalized density profile, defined as  
\begin{equation}
\Delta \rho^\gamma(x) = \frac{ \rho^{\gamma}(x) -
              \rho^{\gamma^*}(x) }{\rho^\gamma_\infty} \; ,
\end{equation}
where $\rho^\gamma(x)$ is the mean density of species $\gamma$ in cells with coordinate $x$,
and $\rho^{\gamma}_{\infty} = 2mn$ is the value of $\rho^\gamma(x)$ far from the interface.

\begin{figure}[h]
\centerline{\includegraphics[scale=0.26,angle=90]{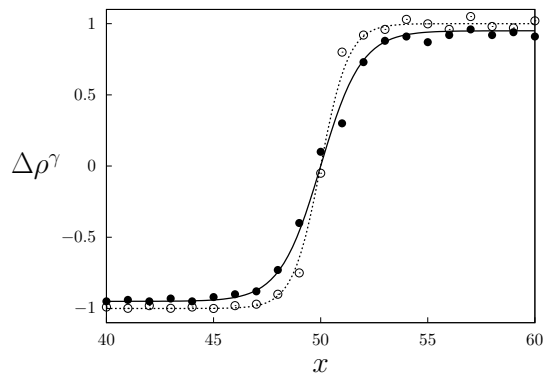}}
\caption{Variation of the normalized profile
density $\Delta \rho^\gamma(x)$ as a function of $x$, for  $n=5$ (open circles)
and $n=8$ (solid circles), with fixed parameter values $T=0.12$,
$\kappa=5.0$. The solid and dashed lines are the fittings of Eq.~(\ref{eq_gx})
for $n=5$ and $n=8$, respectively.}
\label{fig:gx}
\end{figure}

Figure \ref{fig:gx} shows the variation of the normalized density profile for two different number of particles per cell.
Note that, as the number of particles per cell increases, the interface gets sharper. 
This effect is expected since the repulsion between particles of different
species increases with the increment of their respective densities. 

\begin{figure}[h] 
\centerline{\includegraphics[scale=0.28,angle=90]{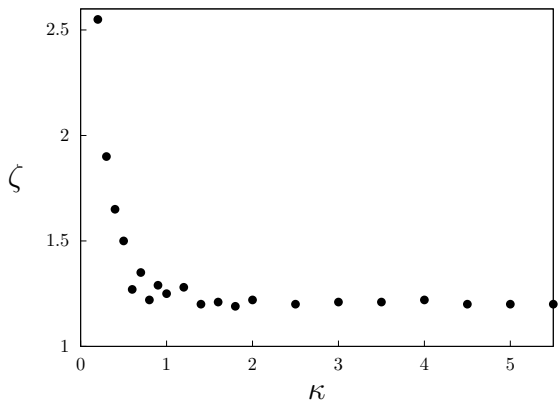}}
\caption{Mean interface width $\zeta$ as a function
of the repulsion parameter $\kappa$ with fixed parameter values $n = 5$ and
$T=0.12$,  after $t=10^5$ iterations.}
\label{fig:zeta_k}
\end{figure}

To measure the interface width, denoted by $\zeta$,
we have fitted the simulation points in the interface profile of Fig.~\ref{fig:gx} with the function \cite{TSafran},
\begin{equation}
\label{eq_gx}
\Delta \rho^\gamma(x) = \tanh{\left( \frac{x}{\zeta}\right)} .
\end{equation}
The behavior of the  interface width $\zeta$ as function of the repulsion parameter $\kappa$ is shown in Fig.~\ref{fig:zeta_k}. 
There is a critical value of $\kappa_c \approx 0.6$, above which $\zeta$ reaches an 
asymptotic minimum value. This indicates that repulsion between species due to
collisions, represented in Eq.~(\ref{eq4}), becomes maximum for values $\kappa > \kappa_c$.

The effect of the temperature on the interface width can be
approximated by using the classical Ising model for an interface between two species \cite{TSafran},
\begin{equation}\label{ec:wi}
\zeta \sim \frac{1}{(T_c-T)^{1/2}} \;,
\end{equation}
where $T_c$ is the critical temperature for the formation of the interface.

\begin{figure}[h]
\centerline{\includegraphics[width=0.3\textwidth,angle=90]{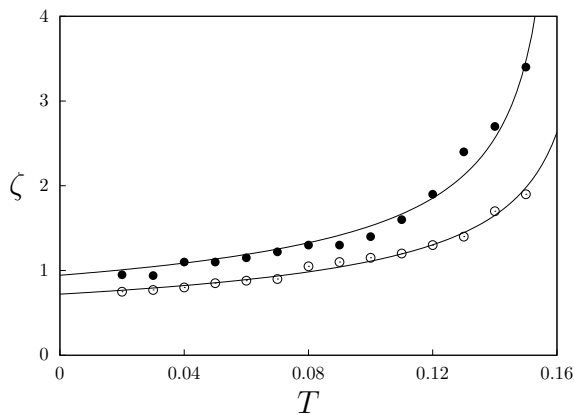}}
\caption{Mean interface width $\zeta$ as a function of
the temperature $T$,  for $n = 5$ (solid circles) and
$n = 8$ (open circles) at $t=10^5$ iterations. 
Fixed parameter $\kappa=5$. Continuous lines correspond to Eq.~(\ref{ec:wi}).} 
\label{fig:WI} 
\end{figure}

In Fig.~\ref{fig:WI} we compare the numerical results obtained from our
model with the values given by Eq.~(\ref{ec:wi}), for two different values of $n$. There is good
agreement between the behavior described by Eq.~(\ref{ec:wi}) and the simulations.
The critical temperatures, calculated from the fitting of the numerical points
to Eq.~(\ref{ec:wi}), are $T_c \approx 0.162$ for $n=5$ and $T_c \approx 0.173$ for $n=8$.

\begin{figure}[h]
\centerline{\includegraphics[scale=0.28,angle=90]{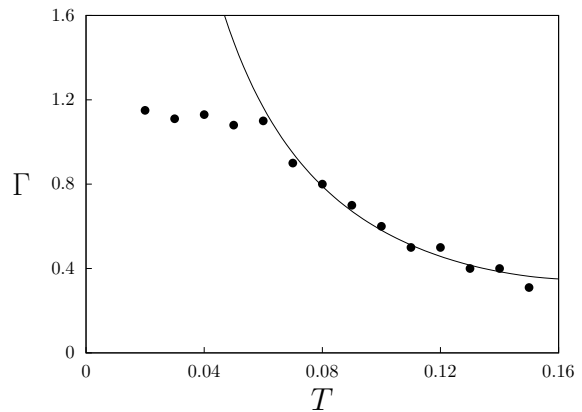}}
\caption{Interface tension $\Gamma$ as a
function of the temperature $T$ at $t=10^4$ iterations.  Fixed parameter values are $\rho^{\gamma}_0 = 5$,
$\kappa=5$.}
\label{fig:TS} 
\end{figure}

Another property that characterizes an interface is the 
interfacial tension, denoted by $\Gamma$. A simple way to estimate $\Gamma$ is through the expression \cite{TSafran}
\begin{equation}\label{ec:TS2}
\Gamma \sim \frac{4\,\rho^\gamma_{\infty}}{3\,\zeta}.
\end{equation}
Figure~\ref{fig:TS} shows $\Gamma$, obtained from Eq.~(\ref{ec:TS2}), as a function
of temperature. The simulation points are compared with the theoretical
expression for interfacial tension given by the Ising model \cite{TSafran},
\begin{equation}\label{ec:TS}
\Gamma \sim \frac{(T_c-T)^{3/2}}{T} .
\end{equation}

Note that, for temperatures $T \geq 0.06$, the interfacial tension
calculated from  the simulations and Eq.~(\ref{ec:TS2}) are well fitted by
the the theoretical curve, Eq.~(\ref{ec:TS}).

It is known, from direct molecular dynamics simulations with two immiscible 
Lennard-Jones fluids \cite{DARF1}, as well as from density functional theory \cite{IF1},
that  the interfacial tension $\Gamma$
exhibits a maximum as the the temperature is varied.
The maximum value of $\Gamma$ arises at a temperature such the attractive interaction forces
between particles cancel out the thermal effect. For temperatures above
this point, the thermal effect is sufficiently strong to cause a decrease of the value of $\Gamma$.
Our multiparticle collision binary fluid model agrees well with the behavior of $\Gamma$ predicted by
the Ising model for temperatures at which the thermal effect is dominant. However,
in the multiparticle collision binary fluid model, the attractive interaction forces between particles are not explicit, 
but they are represented by rotation operators.

Next, we consider the problem of phase separation of an immiscible
binary fluid in the framework of our model. The dimensions of the sides of the box 
are $L_x=L_y=100$, and $L_z=2$, and we assume that the volume has periodic boundary conditions in all three axes.
We start from homogeneous initial conditions where both species 
are uniformly distributed in the volume of the simulation box.

\begin{figure}[h]
\centerline{\includegraphics[scale=0.34,angle=90]{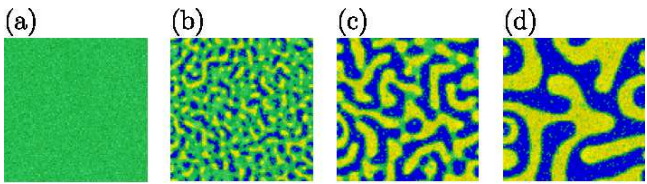}}
\caption{Snapshots along the $z$-axis of the patterns during the evolution of the system, 
with fixed parameter values  $L_x=L_y=100$, $L_z=2$, $\rho^\gamma_0=5$, $T=0.09$ and
$\kappa=5.0$. 
Particles of species $A$ are assigned a yellow (light gray) color and  particles of species $B$ are marked in blue (black). 
Green (dark gray) color indicates
the presence of particles of both species. The color intensity is proportional to the density of particles. 
a) Initial condition. b) $t=10^2$. c) $t=10^3$. d) $t=10^4$.}
\label{fig:formation} 
\end{figure}

Figure~\ref{fig:formation} shows four snapshots of the evolution of the
system. The initial well-mixed state is displayed in Fig.~\ref{fig:formation}(a).
Figures~\ref{fig:formation}(b)-(d), for successive times, show the spontaneous formation of 
of a segregated state, where domains become separated by a thin interface.

A domain growth can be characterized by the time evolution of the average radius, as 
\begin{equation}
 R_t \sim t^\alpha \;,
 \label{ec:R_t}
\end{equation}
where $R_t$ is the average radius of the phase domain at time $t$, and $\alpha$ is the growth exponent. 
We define the average radius $R_t$ as the distance $r$ where 
the spatial correlation function $C[r,t]$ first becomes zero; that is,
\begin{equation}
 R_t = \min \{ r\,\forall\,C[r,t]\} = 0 .
 \label{ec:R_min}
\end{equation}
The spatial correlation function using the discrete cells of the model can be calculated as
\begin{equation}
 C[r,t] = \langle \Phi_t(\xi)  \Phi_t(\xi')\rangle_{\xi,\xi'} ,
 \label{ec:Correltion}
\end{equation}
where $r = |{\bf r}_\xi - {\bf r}_{\xi'}|$ is the distance between the
center of cell $\xi$ and the center of cell $\xi'$, $\langle \cdots \rangle_{\xi,\xi'}$ is
the spatial average over all pairs of cells $\xi$ and $\xi'$ separated a distance $r$, and
\begin{equation}
\Phi_t(\xi) = \rho^\gamma_t(\xi) - \rho^{\gamma^*}_t(\xi) 
\end{equation}
is the difference between the densities of species $\gamma$ and $\gamma^*$ in the cell $\xi$ at time $t$.

\begin{figure}[h]
\centerline{\includegraphics[width=0.6\linewidth,angle=90]{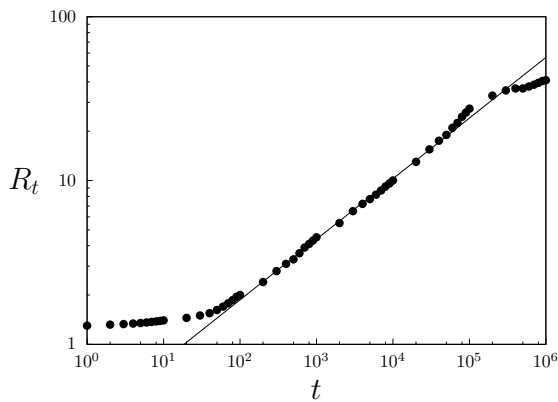}}
\caption{Average radius of a domain phase $R_t$, shown as black dots, as a function of
time $t$ in log-log scale. Fixed parameter values are $L_x=L_y=200$, $L_z=2$,
$\rho_0^\gamma=5$, $T=0.09$ and $\kappa=5.0$. The continuous line corresponds to the best
fitting of Eq.~(\ref{ec:R_t}) for points in the interval $t \in[10^2,10^5]$.}
\label{fig:R_t}
\end{figure}

Figure~\ref{fig:R_t} shows $R_t$ as a function of time, in a log-log plot, with a fixed temperature.
The logarithm of the radius $R_t$ increases linearly with the logarithm of time
in the interval $t \in [10^2, 10^5]$. The corresponding slope, obtained by fitting
of the data, yields the growth exponent $\alpha \approx 0.37$, which is close to the theoretical value for phase growth
in a diffusive regime \cite{TSafran}. For times greater than $t=10^5$, 
the domain size reaches  half of the size of the simulation box; that is,
$R_t \approx L_ x/4$, and the domain growth slows down.

\begin{figure}[h]
\centerline{\includegraphics[width=0.6\linewidth,angle=90]{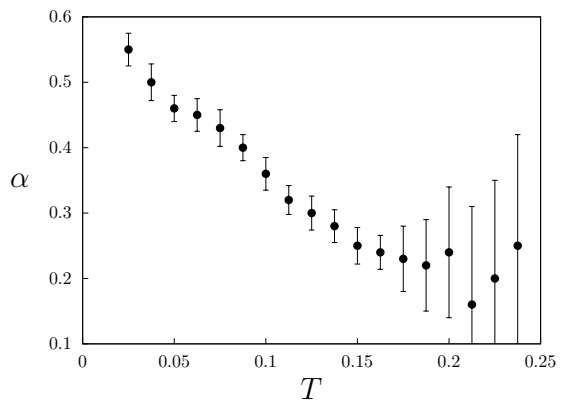}}
\caption{Growth exponent $\alpha$ as a function of temperature
$T$. Fixed parameter values are $L_x=L_y=100$, $L_z=2$, $\rho=5$ and $\kappa=5.0$. Error
bars represent the standard error of the fit of Eq.~(\ref{ec:R_t}).}
\label{fig:alfa_T} 
\end{figure}

Figure \ref{fig:alfa_T} shows the growth exponent $\alpha$ as a function of the 
temperature $T$, calculated numerically from 
data in the time interval for which Eq.~(\ref{ec:R_t}) is valid. 
The exponent $\alpha$ decays linearly with increasing temperature up to a value
$T \approx 0.2$. Above this critical temperature, the error bars in the determination of the quantity $\alpha$ are too large, and
the interface becomes unstable because the thermal mixing destroys the phase separation process.

\section{Binary fluid in crowded environment}
The motion of fluids in crowded environment by obstacles is a problem of much interest in cell biology and other contexts
\cite{Zhou,Verkman,Loren}. 
One way of modeling a film of fluid in a crowded environment is by placing a set
of cylindrical obstacles in the system \cite{EK1}. To simulate the behavior of a binary fluid in a crowded environment with our model,
we insert $N_S$ stationary cylinders in a volume
of radius $\sigma$ and height equal to the height of the simulation box $L_z$.
The fraction of volume occupied by the obstacles is $\phi=2\pi N_S \sigma^2 L_z /{\cal V}$.
We fix the radius of the cylinders at the value $\sigma=2.5$ in units of cells.

As in the previous simulations, particles of one species are distributed uniformly in the
right half-side of the simulation box while particles of the other species are
distributed on the left half-side of the box. We set the particles velocities using a
Boltzmann distribution with temperature $T$, and fix the densities 
$\rho=\rho^\gamma = \rho^{\gamma*}=5$ and the repulsion parameter at the value 
$\kappa=5$. We assume that, when a particle of either species collides with an obstacle, its velocity
is reversed, i.e., a bounce back collision occurs.

\begin{figure}[h]
\centerline{\includegraphics[width=0.43\linewidth,angle=90]{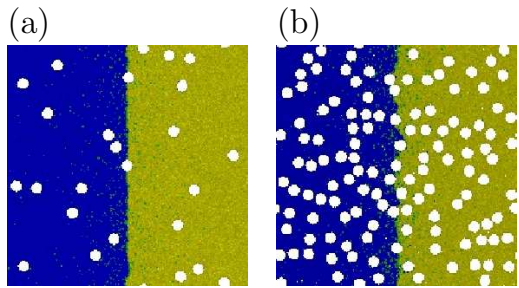}}
\caption{Snapshots along the $z$-axis of the system forming a stable interface at $t=10^4$ iterations,
for different values of the volume fraction of obstacles $\phi$.  White circles indicate the cylindrical obstacles.
Fixed parameters are $\rho = 5$, $\kappa = 5.0$ and
$T=0.06$. (a) $\phi = 0.05$. (b) $\phi = 0.25$. }
\label{fig:IfaseCrow} 
\end{figure}

Figure~\ref{fig:IfaseCrow} shows two snapshots of the system when the interface has stabilized, 
for different values of the fraction of volume occupied by obstacles.
Note that the interface lies close to  obstacles and looks distorted.
This effect is consequence of the reduction of pressure
that occurs when the distance between the interface curve and an obstacle is
small enough to produce an imbalance of forces that removes the particles
lying between the interface and the obstacle. 

We also study the phenomenon of spontaneous phase separation of an immiscible binary fluid in a crowded environment. 
In this case, the particles of each
species are initially distributed uniformly throughout the volume of the simulation
box, avoiding the space occupied by the obstacles.

\begin{figure}[h]
\centerline{\includegraphics[scale=0.3,angle=90]{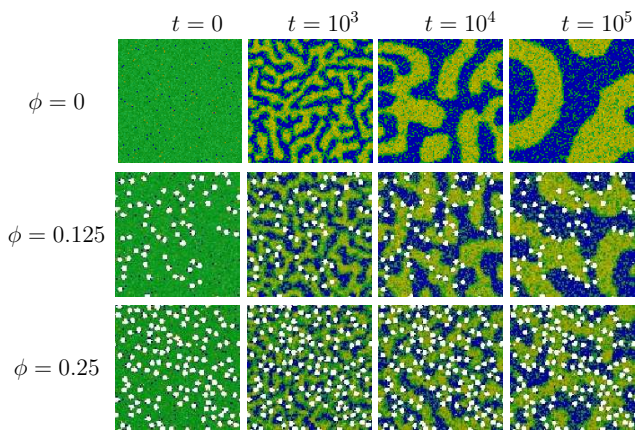}}
\caption{Snapshots along the $z$-axis of the evolution of the phase growth for
three different values of the volume fraction, $\phi=0.0$, $0.125$ and $0.25$
(from top to bottom) and for four different times $t=0$, $10^3$,
$10^4$ and $10^5$ (from left to right).  White circles indicate the cylindrical obstacles.
Fixed parameter values are  $\rho^\gamma = \rho^{\gamma*} = 5$, $\kappa = 5.0$ and
$T=0.09$.}
\label{fig:IphaseCrow} 
\end{figure}

Figure~\ref{fig:IphaseCrow} shows the snapshots of the evolution of phase growth processes
for three different values of the volume fraction of obstacles $\phi$.
As expected, the phase growth process from a homogeneous state is
affected by the presence of obstacles. At first glance, we can see that growth becomes slower when
$\phi$ is increased, since obstacles impede the aggregation of
particles in domains.  
Additionally, we see that the interfaces lie along the locations of obstacles, lacking the rounded profile 
that they possess when the media is free.

\begin{figure}[h]
\centerline{\includegraphics[width=0.33\linewidth,angle=90]{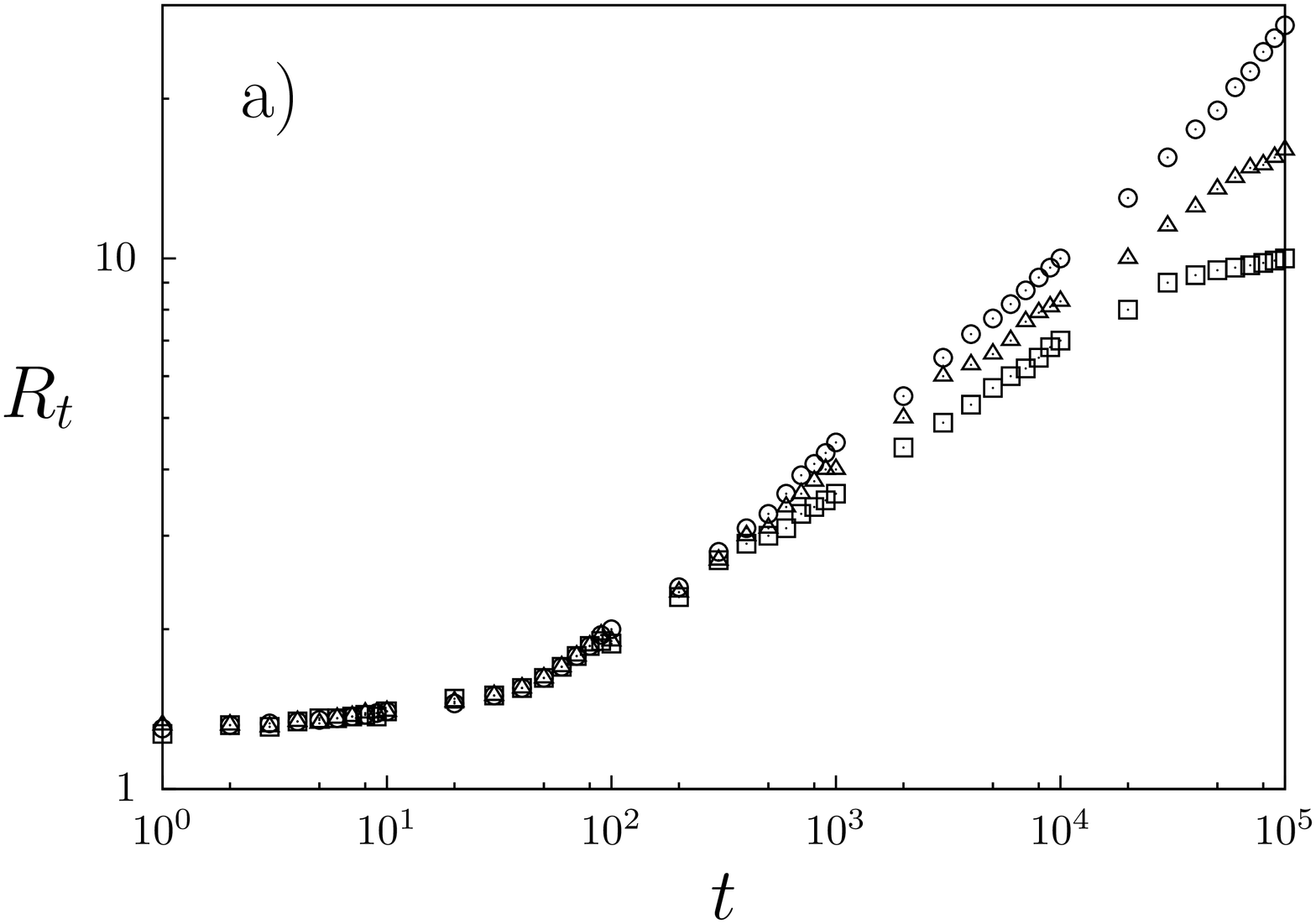}
\hspace{0.5cm}
\includegraphics[width=0.33\linewidth,angle=90]{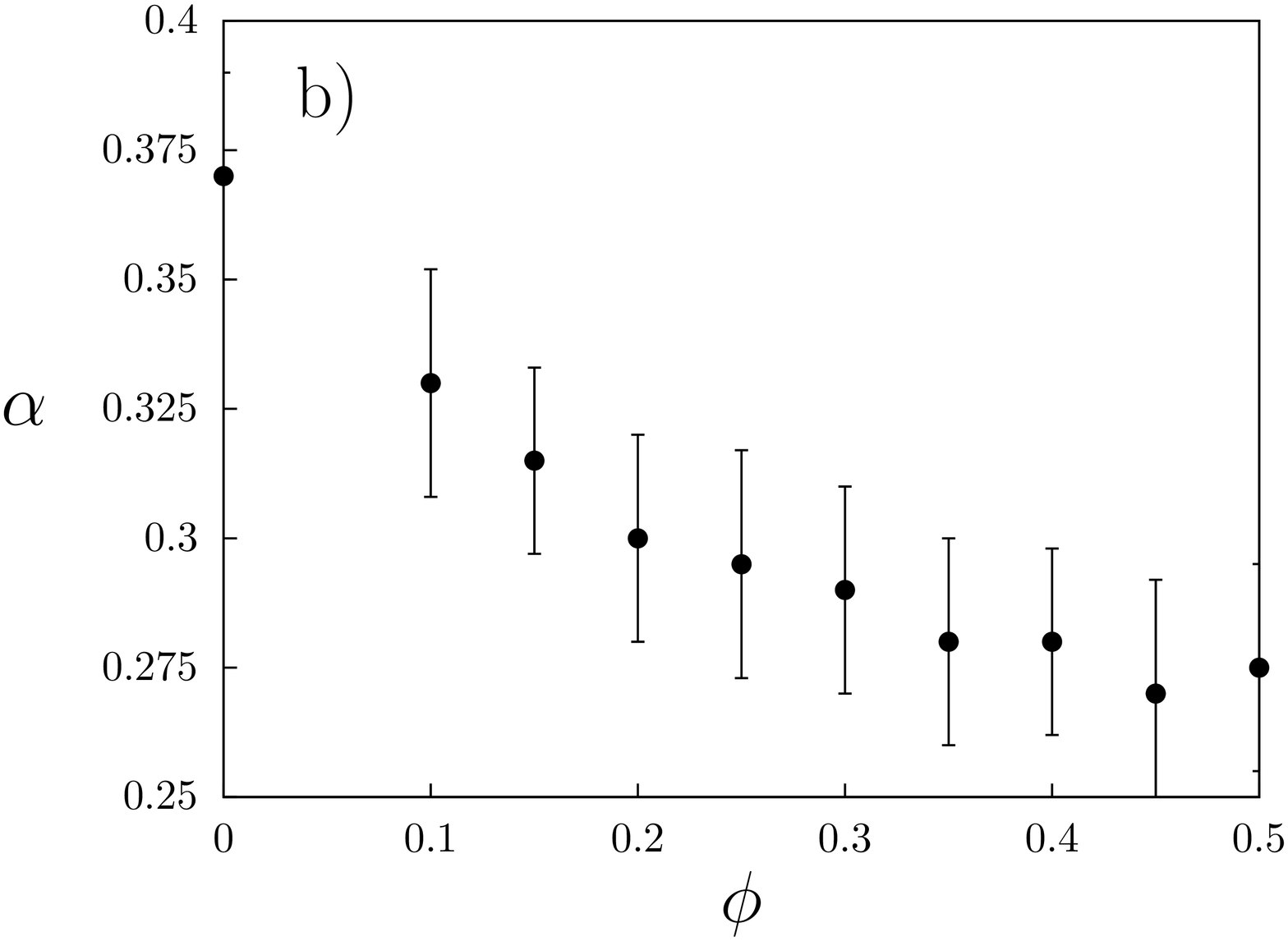}}
\caption{(a) Average radius of the phase domain $R_t$ as function of time $t$ in the presence of obstacles, in log-log scale for $\phi= 0.0$
({\Large $\circ$}), $\phi=0.125$ ($\triangle$) and $\phi=0.25$ ($\square$).
(b) Growth exponent $\alpha$ as function of $\phi$ in the presence of obstacles.
Fixed parameter values are $L_x = L_y = 50$, $L_z = 2$, $\rho^\gamma = \rho^{\gamma*} = 5$, $T = 0.09$ and $\kappa = 5.0$.
Error bars represent the standard error of the fit of Eq.~(\ref{ec:R_t}).}
\label{fig:alfa_phi} 
\end{figure}

To investigate the effect of obstacles on the phase growth process, 
we have calculated the time evolution of the
average radius of the phase domain $R_t$ for several values of
$\phi$, as shown in Fig.~\ref{fig:alfa_phi}(a). For each value of $\phi$, 
$R_t$ can be adjusted to the expression
Eq.~(\ref{ec:R_t}) in the time interval $t \in [10^2,10^5]$. This allows to calculate
the growth exponent $\alpha$ as a function
of $\phi$, as shown in Fig.~\ref{fig:alfa_phi}(b). We can see that increasing the density of obstacles leads to a decrease in
the velocity of the phase growth process, represented by the exponent $\alpha$.

\section{Conclusions}
We have proposed a multiparticle collision dynamics model to investigate phase separation processes in a binary fluid. 
To this aim, we have introduced a repulsion rule between the centers of mass of particles from different species to simulate segregation in
a binary fluid. We have applied this model to mesoscopic systems in both free and crowded environments, where the volume has been discretized 
into finite cells.

Since the repulsion rule only depends on the configuration of the particles inside each cell, 
the model maintains Galilean invariance. In addition, the multiparticle-collision model with repulsion
conserves the mass and the energy of the system
at micro and macro scales, while momentum is conserved within 
homogeneous domains, but not at the interfaces. 

In spite of this limitation, the multiparticle-collision repulsion model yields results consistent
with the known behavior of binary fluids. Properties such as diffusion coefficient, density profile and
width of the interface, calculated from simulations of the multiparticle-collision repulsion model, 
agree very well with the theoretical values predicted by the Ising model for interfaces in a wide range of
temperatures and densities. 
For moderately and low temperatures, the model is able to simulate the
segregation of an immiscible binary fluid into domains, starting from 
mixed, homogeneous initial conditions.
Moreover, the growth exponents for the phases obtained from the model are similar to the
corresponding growth exponents that characterize Newtonian binary fluids.

We have extended the multiparticle-collision repulsion model to simulate crowded environments.
The results from the simulations are also consistent with the behavior of
binary fluids in these environments. 

The good performance of the multiparticle-collision repulsion model for a binary fluid suggests that it can be generalized 
to  incorporate other phenomena, such as  chemical reactions among the species, and to consider species that may diffuse at different rates.
In addition, because of its low computational cost, the multiparticle-collision repulsion model
can be used to simulate systems with relatively large scales of time
and space; i.e., simulate systems with millions of particles per millions
of iterations.

\section*{Acknowledgments} 
This work was supported in part by project No. C-1906-14-05-B from 
Consejo de Desarrollo Cient\'ifico, Human\'istico, Tecnol\'ogico y de las Artes,
Universidad de Los Andes, M\'erida, Venezuela. 
M.G.C. is grateful to the Associates Program
of the Abdus Salam International Centre for Theoretical Physics, Trieste, Italy, for visiting opportunities.

\end{document}